# M-center in low-energy electron irradiated 4*H*-SiC

T. Knežević,[1] A. Hadžipašić,[1,2] T. Ohshima,[3] T. Makino,[3] and I. Capan[1,a]

[1] *Ruđer Bošković Institute, Bijenička 54, 10 000 Zagreb, Croatia*

[2] *Faculty of Science, University of Sarajevo, Zmaja od Bosne 33-35, 71 000 Sarajevo, Bosnia and Herzegovina*

[3] *National Institutes for Quantum Science and Technology, 1233 Watanuki, Takasaki 370-1292, Japan*

We report on the low-energy electron irradiated 4*H*-SiC material studied by means of deep-level transient spectroscopy (DLTS) and Laplace-DLTS. Electron irradiation has introduced the following deep level defects: $EH_1$ and $EH_3$ previously assigned to carbon interstitial-related defects, and M-center, a metastable defect also recently assigned to carbon interstitial defects. We propose that $EH_1$ and $EH_3$ are identical to $M_1$ and $M_3$ and assign them to $C_i^=(h)$ and $C_i^0(h)$, respectively. Moreover, we provide direct evidence that Laplace-DLTS can be used as an excellent tool to distinguish otherwise identical DLTS signals associated with $S_1$ ($V_{Si}$) and $EH_1$ ($C_i$).

Electrically active defects in n-type 4*H*-SiC have been studied in detail for decades. Piece by piece, the puzzle behind the most common and dominant defect traps such as $V_C(=/0)$, $V_C(0/++)$, and $V_{Si}$ has been solved. Part of the puzzle that has kept researchers busy for several years is the study of silicon vacancy and carbon interstitial defects ($V_{Si}$ and $C_i$), introduced by radiation. It was well known that electron irradiation[1-3], proton irradiation[4-6], neutron irradiation[7,8], and ion implantation[9,10] introduce two deep level defects in n-type 4*H*-SiC material. These levels are located at 0.40 and 0.70 eV below the conduction band and have been labeled either as $S_{1/2}$ or $EH_{1/3}$. Recently, Bathen *et al.*[6] have provided conclusive evidence that $S_{1/2}$ deep level defects are related to $V_{Si}$, while Alfieri *et al.*[3] have shown that $EH_{1/3}$ deep level defects are related to $C_i$. The $EH_{1/3}$ deep level defects are introduced only in the case of the low-energy electron irradiation (< 200 keV), since under such conditions silicon atoms cannot be displaced[1,2].

The $V_{Si}$ has attracted much attention in recent years due to its physical properties and its potential application for quantum sensing[6,11-13]. The $S_1$ and $S_2$ are identified as $V_{Si}$ (−3/=) and $V_{Si}$ (=/−) charge transitions, respectively. Bathen *et al.*[6] have shown that $S_1$ (in proton irradiated 4*H*-SiC samples) has two emission lines originating from $V_{Si}$ sitting at -*k* and -*h* lattice sites. These findings were later confirmed by Capan *et al.*[8] when studying 4*H*-SiC material irradiated with fast neutrons.

Despite their technological importance, $C_i$ defects are not yet fully understood. Coutinho *et al.*[14] have recently shown that a bi-stable defect in 4$H$-SiC, known as M-center, is carbon interstitial. Accordingly, the defect is responsible for two pairs of first and second acceptor transitions[4,5,10,14,15],

$$\text{Configuration A} \rightarrow \{M_1(=/-) = E_c - 0.42 \text{ eV}; M_3(-/0) = E_c - 0.74 \text{ eV}\}$$

$$\text{Configuration B} \rightarrow \{M_2(=/-) = E_c - 0.65 \text{ eV}; M_4(-/0) = E_c - 0.86 \text{ eV}\}$$

where configurations A and B were assigned to a carbon interstitial at the hexagonal and cubic sub-lattice sites, $C_i(h)$ and $C_i(k)$, respectively. The two configurations can be interchanged by annealing and applying a reverse bias voltage. Configuration A is obtained when the measurement is performed after annealing the sample just above room temperature under reverse bias voltage. The defect jumps to configuration B after the sample is annealed at ~450 K without bias.

The evident similarities between $M_{1/3}$ and $EH_{1/3}$ traps, including their location within the band gap and their formation conditions, are so striking that we must hypothesize that they may ultimately arise from the same defect. Therefore, the main goal of this work is twofold. By using low energy electrons and low fluence, we intend to introduce only the $EH_{1/3}$ deep level defects and verify if there is evidence for bi-stability and the formation of $M_{1/3}$. Moreover, since we displace only the carbon atoms, resulting in a very clear $EH_1$ signal, Laplace-DLTS was used to investigate possible superposition of $EH_1$ and $M_1$ signals.

In this work, n-type Schottky barrier diodes (SBDs) were fabricated on nitrogen-doped (~ 4.7×10$^{14}$ cm$^{-3}$) 4$H$-SiC epitaxial layers, with a thickness of approximately 25 μm. The epi-layer was grown on an 8° off-cut silicon face of a 350 μm thick 4$H$-SiC (0001) wafer without a buffer layer by chemical vapor deposition. The Schottky barriers were formed by thermal evaporation of nickel through a metal mask with a patterned squared aperture of 1 mm edge length, while the Ohmic contacts were formed on the backside of the silicon carbide substrate by nickel sintering at 950 °C in an Ar atmosphere.

Low-energy electron irradiations were performed at NHV, Kyoto, Japan. The electron energy was 150 keV and the total fluence was 1×10$^{15}$ cm$^{-2}$.

DLTS measurements were performed using a Boonton 7200 capacitance meter (Boonton Electronics, New Jersey, USA) and an NI PCI-6521 data acquisition device (NI, Austin, USA). Conventional DLTS measurements were carried out in the temperature range of 100 to 450 K with a temperature ramp rate of 2 K/min. Reverse voltage, pulse voltage, and pulse width were $V_R$=−4V, $V_P$=0V, and $t_P$=10ms, respectively. For the Laplace-DLTS measurements the following acquisition settings were used: the number of samples $3 \times 10^4$, the sampling rate 10–80 kHz, and the number of averaged scans 100–800. The numerical routine FLOG[16] was used to calculate Laplace-DLTS spectra.

M-center was transformed to configuration B and configuration A by annealing at 450 K (for 20 min) and cooling the SBD without applying a bias voltage (0 V), or by annealing at 340 K for 20 min and cooling down the SBD with an applied bias voltage of −30 V, respectively.

Figure 1 shows the DLTS spectrum for the n-type 4$H$-SiC SBD irradiated with low-energy electrons in configurations A and B. It should be noted that the DLTS spectrum for the as-grown 4$H$-SiC SBD is not shown here, as it has been already reported in several previous studies[17,18]. In the as-grown sample, only the $Z_{1/2}$ peak is present. The $Z_{1/2}$ has already been assigned to Vc(=/0)[19] .

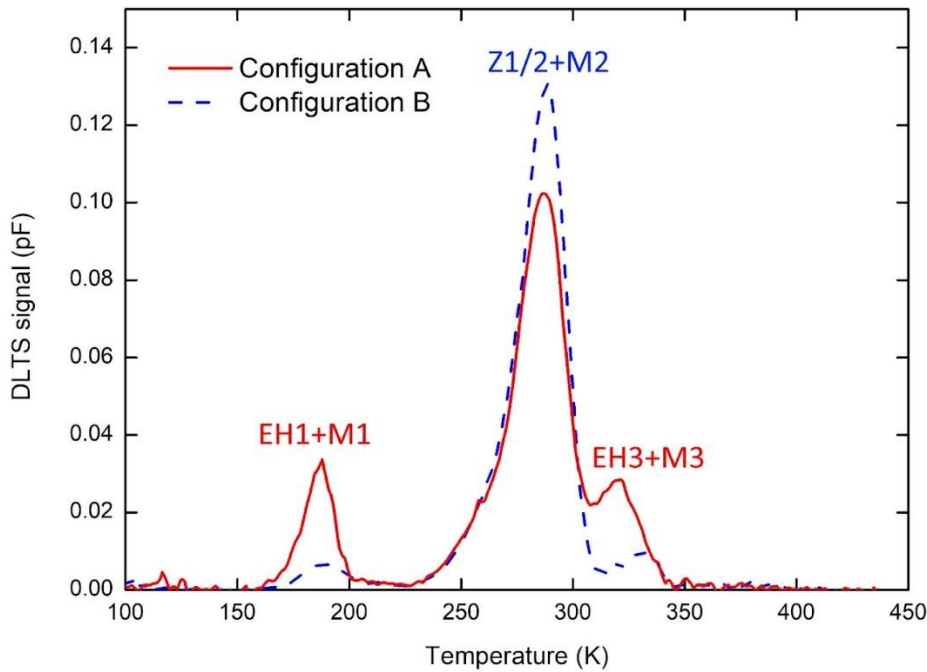

FIG. 1. DLTS spectra for the 4$H$-SiC sample irradiated with low-energy electrons, measured in configurations A and B.

Irradiation with low-energy electrons introduced two deep-level defects, whose positions (0.41 and 0.70 eV) are close to those that are usually labeled as $EH_1$ and $EH_3$. However, the observed peaks are also metastable and consistent with the properties of the M-center[4,5,10,14,15]. The spectra depend on the applied bias voltage and annealing. The M-center introduces four deep-level defects. $M_1$ and $M_3$ overlap with $EH_1$ and $EH_3$ in configuration A, while $M_2$ overlaps with $Z_{1/2}$ in configuration B (Fig.1). M-states and their bi-stability are more clearly observed if the DLTS signal difference (configuration A – configuration B) is plotted, as shown in Figure 2. As previously reported, it is not possible to observe the $M_4$ with DLTS due to the technical limitations, but $M_4$ has been observed in ion-implanted[10] and neutron-irradiated[8] $4H$-SiC using the isothermal DLTS.

We have estimated the activation energies for $M_1$, $M_2$, and $M_3$ and obtained the following values 0.42, 0.73, and 0.90 eV, which are consistent with previously reported values[4,5,10,14,15].

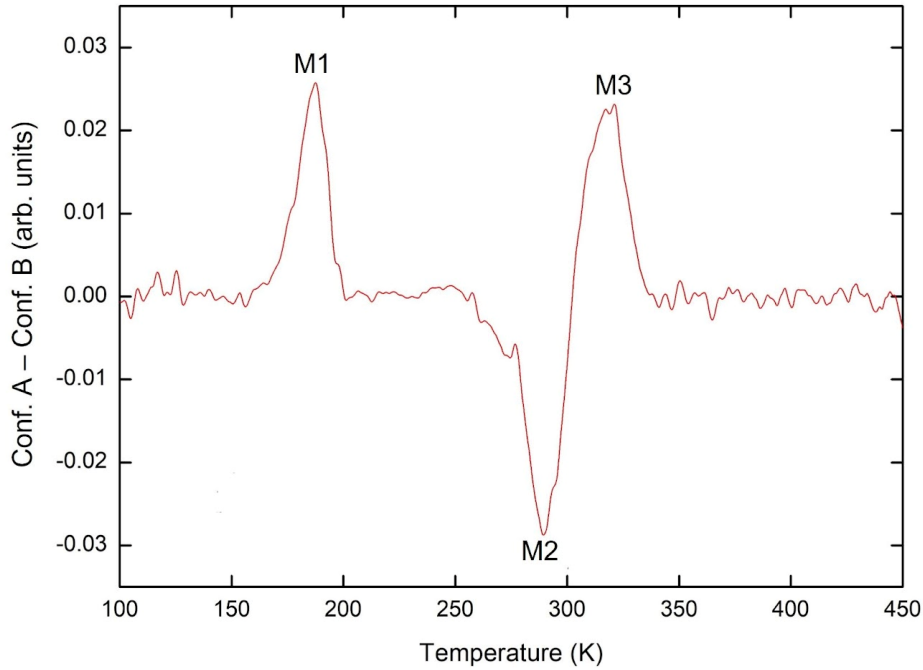

FIG. 2. The DLTS signal difference (configuration A – configuration B) for the $4H$-SiC sample irradiated with low-energy electrons.

The DLTS data obtained in this study are in perfect agreement with previous results from DLTS studies on low-energy electron irradiated 4H-SiC samples[1]. However, triggered by the recent advancements in understanding and identifying the M-center[14], accompanied by the recent progress in understanding the $EH_{1/3}$ and $S_{1/2}$ defects[3,6], we have applied the Laplace-DLTS technique to study the $EH_1$ in more details and to investigate the correlation between the EHs defects and the M-states.

Figure 3 shows Laplace-DLTS spectra in configurations A and B, measured at the temperature of the $EH_1$ DLTS peak (T = 200 K). The Laplace-DLTS results clearly indicate that $EH_1$ is a single defect, with no splitting of the emission line due to the different lattice sites (-$h$ and -$k$) as is the case for $S_1$ and/or $Z_{1/2}$ ($EH_{6/7}$)[6,8,17,18,20]. Thus, our results clearly show that Laplace DLTS can be used as a tool to distinguish $S_1$ and $EH_1$, which give identical signals in the DLTS spectrum. To our knowledge, this is the first time that $EH_1$ has been directly measured with Laplace-DLTS.

According to the recent findings of Coutinho *et al.*[14], the M-center was assigned to $C_i$. The four states arising from the M-center are assigned to different charge states located at the different lattice sites. $M_1$ and $M_3$ are assigned to $C_i^=(h)$ and $C_i^0(h)$, while $M_2$ and $M_4$ are assigned to $C_i^=(k)$ and $C_i^0(k)$. The occupancy of the *-h* or *-k* lattice sites is controlled by the reverse bias voltage anneals. For example, if we perform zero bias annealing (configuration B), the *-k* sites will be dominantly occupied, leading to the appearance of $M_2$ and $M_4$ in the DLTS spectra. However, if we perform reverse bias annealing (configuration A), then the occupation of the *-h* sites prevail, which gives rise to $M_1$ and $M_3$ in the DLTS spectra.

According to the Laplace-DLTS measurements at temperatures around 200 K (i.e. temperature at which $EH_1$ has the peak maximum in the DLTS spectrum), only one emission line is observed in both configurations, A and B. There is no convincing evidence or even suggestions for two overlapping defects with identical emission lines, that are unresolved by Laplace-DLTS. Let us assume that $M_1$ is indeed $EH_1$. The difference in the intensity of $EH_1$ peak (DLTS spectra) in configurations A and B is not due to the overlap of an additional defect such as $M_1$, but to the different occupancy of $C_i^=(h)$ sites. For configuration A, as explained above, this is more favorable than *-k* sites. Therefore, we can speculate that $EH_1$ is the same defect as $M_1$, and assigned to $C_i^=(h)$.

Based on the difference signal (configuration A – configuration B), as shown in Fig.2, the concentrations of $M_1$ and $M_2$ are identical within the measurements error margin. $M_2$ has been recently assigned to $C_i^=(k)$[7]. This leads us to conclusion that conversion $C_i^=(h) \leftrightarrow C_i^=(k)$ occurs more easily than anticipated. As mentioned above, we control the occupancy of the *-h* and *-k* sites by different bias voltage annealing. From isothermal annealing, the conversion from configuration A to configuration B was previously measured to be activated with a barrier of 1.4 eV, while the conversion from configuration B to configuration A is activated with a lower barrier of 0.9 eV[4]. It should be noted that these values were estimated in the study of MeV proton implanted 4*H*-SiC material. The MeV implantations result in the introduction of the $S_{1/2}$ defects ($V_{Si}$). By varying the filling pulse length while maintaining the measurement temperature at the $S_1/M_1$ peak position, Martin *et al.*[4] have clearly shown that contributions from at least two different defects, $S_1$, and $M_1$ are present in configuration B. These results were recently confirmed by Capan *et al.*[8] as $S_1$ and $M_1$ have directly been measured with Laplace-DLTS. Not only do we have contributions from $S_1$ and $M_1$, but $S_1$ is additionally resolved into two components. Although the analysis of the conversion barriers (configuration A $\leftrightarrow$ configuration B) has been done using the "difference" DLTS signal (presumably this is the case where the signal is only due to the M-center)[4], we should not completely underestimate the fact that the conversion barriers were not determined under conditions equivalent to those reported in this study. The conversion barriers $C_i^=(h) \leftrightarrow C_i^=(k)$ could be lower than previously assumed.

All these results imply that EH$_{1/3}$ and M-center are indeed carbon interstitials, and they are all arising from the same defects. Thus, we conclude that EH$_1$ and EH$_3$ are identical to M1 and M3, and assign them to $C_i^=(h)$ and $C_i^0(h)$, respectively.

Unfortunately, Laplace-DLTS cannot provide conclusive information about EH$_3$ as is the case with EH$_1$, since the EH$_3$ is too close to Z$_1$(=/0) and Z$_2$(=/0) in the Laplace DLTS spectrum, and they overlap. The conversion $C_i^0(h) \leftrightarrow C_i^0(k)$ should follow the same path. Further studies with isothermal DLTS are needed since this is the only way to measure M$_4$ directly.

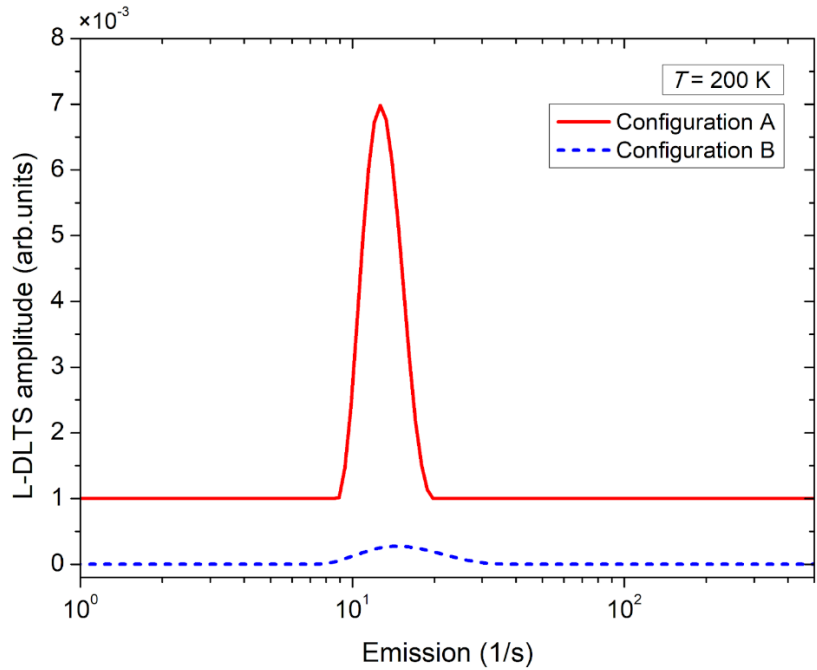

FIG. 3. Laplace DLTS spectra of the 4*H*-SiC samples irradiated with low-energy electrons, measured in configurations A and B at a measurement temperature of 200 K.

In this work, we have provided direct evidence that Laplace-DLTS can be used as an excellent and practical tool to distinguish otherwise identical DLTS signals associated with S$_1$ (V$_{Si}$) and EH$_1$ (C$_i$). These signals have caused much confusion in the labeling and identification of irradiation-induced deep level defects located at 0.40 and 0.70 eV below the conduction band. Moreover, based on the results obtained in this study, we propose that EH$_1$ and EH$_3$ are identical to M$_1$ and M$_3$, and assign them to $C_i^=(h)$ and $C_i^0(h)$, respectively.

## ACKNOWLEDGMENTS

This work was supported by the NATO SPS Programme through Project No. G5674.

## AUTHOR DECLARATIONS

### Conflict of Interest

The authors have no conflicts to disclose.

## DATA AVAILABILITY

The data that support the findings of this study are available from the corresponding authors upon reasonable request.

L180102 (2021).